# The Transformative Potential of Artificial Intelligence

Ross Gruetzemacher[1] and Jess Whittlestone[2]
[1] Wichita State University, W. Frank Barton School of Business, ross.gruetzemacher@wichita.edu
[2] University of Cambridge, Leverhulme Centre for the Future of Intelligence, jlw84@cam.ac.uk

## Abstract

The terms 'human-level artificial intelligence' and 'artificial general intelligence' are widely used to refer to the possibility of advanced artificial intelligence (AI) with potentially extreme impacts on society. These terms are poorly defined and do not necessarily indicate what is most important with respect to future societal impacts. We suggest that the term 'transformative AI' is a helpful alternative, reflecting the possibility that advanced AI systems could have very large impacts on society without reaching human-level cognitive abilities. To be most useful, however, more analysis of what it means for AI to be 'transformative' is needed. In this paper, we propose three different levels on which AI might be said to be transformative, associated with different levels of societal change. We suggest that these distinctions would improve conversations between policy makers and decision makers concerning the mid- to long-term impacts of advances in AI. Further, we feel this would have a positive effect on strategic foresight efforts involving advanced AI, which we expect to illuminate paths to alternative futures. We conclude with a discussion of the benefits of our new framework and by highlighting directions for future work in this area.

# 1. Introduction

> "*AI is one of the most important things we're working on ... as humanity. It's more profound than fire or electricity or any of the bigger things we have worked on. It has tremendous positive sides to it, but, you know it has real negative consequences, [too].*" - Sundar Pichai (Pichai and Schwab 2020)

Artificial Intelligence[1] (AI) has seen dramatic progress in recent years, particularly in the subfield of machine learning known as deep learning. This progress has raised concerns about the potential applications of these advances and their impact on society. These concerns are shared by AI researchers, science and technology policy professionals, as well as the general public (Zhang and Dafoe 2019).

While it is difficult to predict future technological progress, it is plausible that more advanced AI systems could precipitate dramatic societal changes. A principal goal of the field of AI has long been to build a machine with humanlike "common sense" (Minsky et al. 2004), and Turing (1950) famously proposed an 'imitation game' for evaluation of intelligent machine behavior relative to humans. Several different terms have been used to refer to the possibility of these humanlike AI systems with the potential to lead to such changes, including "human-level AI" (HLAI; McCarthy 2007), "high-level machine intelligence" (HLMI; Grace et al. 2018), and "artificial general intelligence" (AGI; Baum et al. 2011). These notions all imply that most of our concern should be afforded to systems which are human-like or sufficiently general in their capabilities. However, not all visionaries who contributed to the birth of the field of AI ascribed anthropomorphic qualities to "general-purposes computers". Herbert Simon notably anticipated such advanced AI as being able to substitute for any human functions in organizations but discussed the capabilities and impacts of such a technology in purely economic terms (Simon 1965).

In recent years, the notion of **transformative AI (TAI)** has begun to receive traction among some scholars (Karnofsky 2016; Dafoe 2018), to reflect the possibility that certain types of advanced AI systems could have transformative effects on society without having human-level cognitive abilities.

We believe that it is good practice to use the term TAI to refer to advanced AI systems with much greater potential for societal impact (rather than, for example, AGI or HLAI), because it captures the idea that a broad spectrum of advanced AI systems are worthy of concern. However, the broad inclusivity of the term TAI is a limitation as well as a strength. The term has been used increasingly in recent years (Author et al. 2020; Trammell and Korinek 2021; Zhang and Dafoe 2019; Dafoe 2018; Horowitz 2018), as have references to AI "transforming" life more colloquially (McWaters 2018; West and Allen 2018; White House 2018), but authors and speakers are often ambiguous in what they consider transformative. This ambiguity limits our ability to understand, anticipate, forecast, and communicate clearly about a range of possible future AI scenarios, which has significant implications for futures researchers and practitioners as well as the AI research community[2]. For this reason, different interpretations of the term 'transformative' in the context of AI should be more clearly delineated.

---

[1] There is no consensus among experts on how to define AI (Author 2020), but we adopt a definition of AI consistent with mainstream thinking: AI does not describe any one specific application, but rather it refers to a set of computational techniques (Stone et al. 2016) with the objective of enabling machines to behave intelligently (Russell and Norvig 2010).

Defining TAI is further complicated by the fact that others have used the notion of 'societal transformation' more informally to refer to impacts that AI is *already* starting to have on society[3] (West and Allen 2018). Leading business consulting firms are widely discussing how AI is beginning to transform business and society (McWaters 2018), and others have discussed (managing) the transformative effects that AI is having on organizations while comparing AI's future impact to that of electricity (Ng 2018). Widespread surveillance is currently possible given sufficient funding and appears to have the potential to irreversibly change the ability of authoritarian governments to suppress dissent (Agarwal 2019; Turchin 2018), but whether this occurs depends more on political factors than technological ones. Still, such transformative effects fall far short of societal impacts "comparable to (or more significant than) the agricultural or industrial revolution" (Karnofsky 2016).

In this article, we analyze the notion of transformative AI, considering different levels of societal transformation that AI could plausibly lead to. We draw from a variety of perspectives to frame the proposed levels in the broad bodies of existing literature on economic history and technology-driven societal change, and we discuss specific avenues of technical progress with potential to lead to each level of transformation.

We intend for this analysis to help clarify conversations in the AI research community around anticipating different types of advanced AI, the potential impacts of different advances, and corresponding research priorities. In doing so we foresee this work having a substantive impact on futures research given the lack of a framework for facilitating constructive discussions amidst the wide scope of plausible futures advanced AI may precipitate (Turchin 2019; Makridakis 2017). For example, the divergent futures presented by Turchin and Makridakis represent but a small portion of the space of all plausible futures that may arise from advanced AI. We hope that the framework presented in this paper will help to shift anticipatory assumptions and in doing so open up a much broader exploration of the space of possible AI futures.

We proceed by first examining the existing definitions of TAI. We then review broad bodies of literature from economic history and technology-driven societal change, and use data found in this process to illustrate the nature of change on the level of the agricultural or industrial revolution. In the next section we build on this literature, proposing three levels on which AI may be thought to be transformative, as well as dimensions for further improving discussions on the transformative potential of AI. We conclude with a discussion of the implications for the AI research community and future research

---

[2] We use this term to refer to the community of academics and professionals engaged in AI research from a broad range of disciplines such as the law, public policy, computer science, etc.

[3] We note that AI has a long history of hype cycles that are more commonly referred to by the periods when interest recedes dubbed AI winters (Russell and Norvig 2010). There have been two previous AI winters, following the original explosion of interest in AI during the 1950s, and then again following the second wave of AI interest in the 1980s. Beginning with Turing (1950), and spanning each of the boom-and-bust cycles, there has been interest in humanlike AI systems such as HLAI and beyond (Good 1963; Vinge 1993; McCarthy 2007). There has been some debate over whether the recent progress in AI should be thought of as an AI summer, with a looming AI winter, or whether this time might be different, and lead to a period when interest in AI and AI research persists for an extended period. This debate is beyond the scope of this article, but we mention it here to note that the framework we propose is predicated on a continued AI summer.

## 2. Existing Definitions of Transformative AI

Four different existing definitions of TAI are given in Table 1, all which are somewhat ambiguous. What would count as *radical* changes to welfare, wealth, or power (Dafoe 2018)? What falls between a narrow task, like playing a video game, and superintelligence (Horowitz 2018)? The two remaining definitions (Zhang and Dafoe 2018; Karnofsky 2016) are more specific in making comparisons to the agricultural and industrial revolutions; however, it is still unclear what it would mean for such advances to precipitate change "comparable to the industrial revolution."

**Table 1:** A comparison of previous definitions of TAI.

| Karnofsky 2016 | "potential future AI that precipitates a transition **comparable to (or more significant than) the agricultural or industrial revolution**" |
|---|---|
| Dafoe 2018 | "advanced AI that could lead to **radical changes in welfare, wealth or power**" |
| Horowitz 2018 | "AI that can go **beyond a narrow task … but falls short of achieving superintelligence**." |
| Zhang & Dafoe 2019 | "advanced AI systems whose long-term impacts may be as **profound as the industrial revolution**" |

Many other powerful players from business to government are using notions of "transformation" to describe AI's societal impact more informally. Deloitte, for example, released a report in 2018 on *How artificial intelligence is transforming the financial ecosystem* (McWaters 2018), and the Brookings Institute similarly published a report entitled *How Artificial Intelligence is Transforming the World* (West and Allen 2018). Even the White House has suggested that "AI is quickly transforming American life and American business" (White House 2018). Leading academics have even suggested AI to be a potentially transformative new technology (Brynjolfsson et al. 2019) with the potential to substantially alter nearly all occupations to some degree (Frank et al. 2019).

## 3. Transformative Societal Change in History

Dramatic changes to societal systems due to technological progress are not unprecedented. In order to think clearly about what it might mean for AI to be 'transformative' it is useful to begin by considering the existing literature on technology-driven transformative societal change in history. While previous definitions of TAI have drawn historical comparisons, they have not engaged with this body of literature which discusses a range of different technological transformations. In particular, this literature suggests that there are multiple different types or *levels* of transformative societal change which may be brought about by new technologies.

**Existing literature on transformative technological change**

Much of the literature on transformative technologies has centered around the notion of **general purpose technologies** (GPTs; Bresnahan and Trajtenberg 1995). We adopt Lipsey et al.'s (2005) definition of a GPT as "a technology that initially has much scope for improvement and eventually comes to be widely used[4], to have many uses, and to have many spillover effects", acknowledging that this is very broad. A commonly used example of a GPT is electricity, which led to substantial changes in daily life and communication, enabling products we take for granted today including light bulbs and telephones. Other examples of GPTs include the steam engine, electric motors, semiconductors and computers. GPTs are thought to drive economic growth across many sectors due to their broad applicability to a wide variety of tasks, eventually leading to notable increases in economic productivity.

However, it might also be appropriate to refer to some technologies as "transformative" in a narrower sense, despite them not being sufficiently general in their application to be classed as a GPT. For example, most would agree that the invention of nuclear weapons had a transformative impact on warfare and international relations, but neither nuclear weapons nor nuclear power are broadly considered to be GPTs. We may therefore want to account for the possibility that some technologies, despite having less pervasive economic uses than GPTs, can be transformative by having an extreme impact on a narrow yet important part of economic, social, or political life.

The case has also been made that the agricultural and industrial revolutions constituted transformative change on a higher level than that of other periods (Bocquet-Appel 2011; Pomeranz 2021; McCloskey 2004). Both of these revolutions constituted extreme and unprecedented changes to human life: a transition from people living as hunter-gatherers to large, settled civilizations; and a transition to mechanized manufacturing and factories, leading to unprecedented population growth and rising quality of life (Morris 2013; Clark 2007). The industrial revolution in particular coincided with clear trajectory changes in metrics of human well-being including measures of physical health, economic well-being, energy capture and technological empowerment. While technological advances such as electricity and the internal combustion engine have had transformative impacts on many aspects of human life and well-being, they have not alone changed the nature of civilization in the same fundamental way the agricultural revolution did, and do not appear to have led to such an extreme change in the metrics of human well-being as did the industrial revolution.

The industrial revolution, which occurred in British society from the mid-18th to mid-19th centuries, has been extensively studied by economic historians. It has received such significant attention because it stands apart from other 'efflorescences' in world history in that it is commonly thought to have been the beginning of self-sustaining and accelerating economic growth that has continued to the present (Goldstone 2002; Mokyr 2016, Allen 2017). Mokyr (2005) explains that we should think of it "as a phase transition in economic history, in which the old parameters no longer hold, and in which the system's dynamics have been unalterably changed". Many economists refer to this specific period as the first industrial revolution in order to distinguish it from further societal transformation driven by subsequent innovations (Schwab 2017; Rifkin 2011; Mokyr and Strotz 1998). We use the term industrial revolution in reference not only to the introduction of the steam power and coal use in Britain, transformative innovations which led the continuing economic growth (*i.e.*, "Modern Economic Growth") that followed (Allen 2009), but to the broader phase transition as well. Thus, understood as such, it is more significant than other technological advances like electricity or the internal combustion engine.

---

[4] We feel it is important to note that this definition is adapted to reflect that GPTs could be susceptible to productivity paradoxes (Solow 1987), i.e., that GPTs are often thought to lead to significant increases in economic productivity only after the initial scope for improvement has been partly exhausted and there have been numerous complementary innovations. Many economists are suggesting AI is currently in the midst of a productivity paradox (Brynjolfsson et al. 2019; Krishnan et al. 2018).

Substantial work has also focused on understanding the effects of transformative technologies as part of broader periods of transformative change or 'long waves' (a.k.a. 'Kondratiev waves'; Schumpeter 1939; Kondratiev 1926): extended periods of rapid economic growth driven by temporal 'clusters' of technological innovations (Ayres 1992). Ayres suggests that society has seen five major technological transformations, each associated with clusters of technologies, with the earliest beginning in 1770 and the most recent beginning in 1983 (see Table 2). He identifies the first two technological transformations as equivalent to the first industrial revolution and the third technological transformation as potentially equivalent to the second industrial revolution. Each of these transformations is closely associated with one, and sometimes multiple GPTs. These periods have also been called technological revolutions (Perez 2003), and we refer to the two most extreme such cases - *i.e.*, the agricultural and industrial revolutions - as **production revolutions** (Grinin et al. 2017).

**Table 2:** Five technological transformations/revolutions and associated GPTs (Lipsey et al. 2005; Perez 2003; Ayres 1992).

|  | Time | Technological Transformations/Revolutions | GPTs |
|---|---|---|---|
| 1st | 1770-1800 | Change from water power to large-scale use of coal | Steam power |
| 2nd | 1825-1850 | Steam power applied to textiles and railroads | Factories, railroads |
| 3rd | 1860-1900 | Steel, mechanized manufacturing, illumination, telephones & motors | Electricity, internal combustion engine |
| 4th | 1930-1950 | Advances in synthetic materials & electronics | Mass production |
| 5th | 1980- | The convergence of computers and telecommunications | Computer, the Internet |

Plotting metrics of human progress and well-being may help to clarify what is implied by societal transformation "comparable to (or more significant than) the agricultural or industrial revolution" (Karnofsky 2016). Figure 1 depicts the average global gross domestic product (GDP) and the average global war making capacity from 2000 BCE to 1990 CE[5] (De Long 1998; Morris 2013). The time period associated with the industrial revolution is shaded to depict the sharp change in gradient for each of the measures associated with it. Clearly, no single technological transformation has had as broad an impact on all these measures as the industrial revolution has.

---

[5] GDP can be interpretted as a suitable long-term measure of human progress for obvious reasons, and Morris (2013) suggests that war making capacity is another suitable measure of long-term human progress.

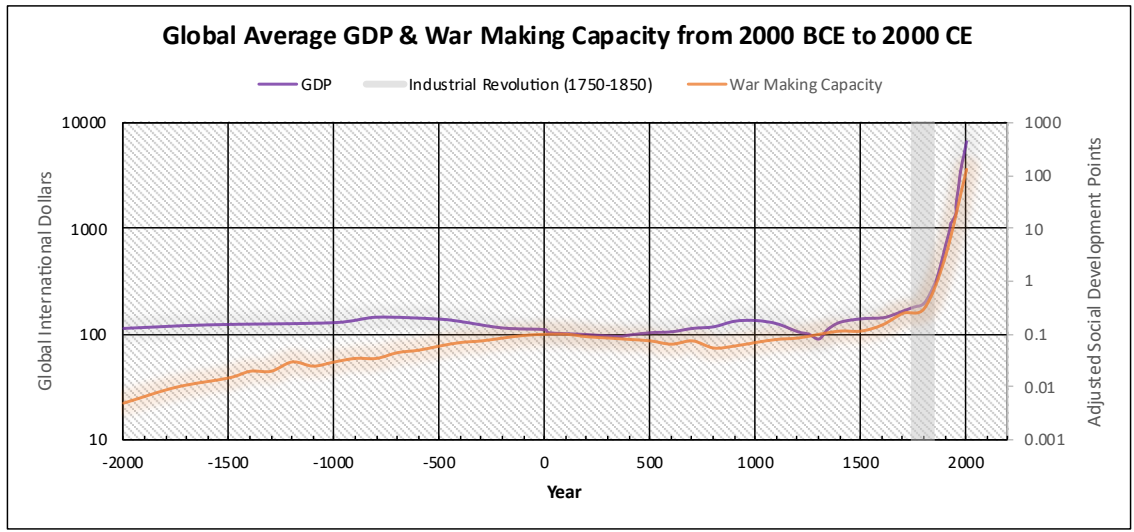

**Figure 1:** This depicts the global average GDP and global average war making capacity from 2000 BCE to 2000 CE (De Long 1988; Morris 2013), with a region shaded in grey to highlight the industrial revolution. Considering that both axes are on a logarithmic scale, the tremendous impact of the industrial revolution on these two measures of human progress is obvious[6]. Changes of this magnitude are unique in human history. Consequently, further transformative change of this magnitude may be difficult to comprehend. To say such change would be extreme could be thought to be an understatement; this plot is intended to convey how radical such change may be.

**Elements of transformative societal change**

This literature on technology-driven societal transformations helps us to elucidate some of the different ways that AI might be said to be 'transformative.' Claiming that AI is well on its way to becoming a GPT – *i.e.*, being applied widely across sectors with many spill-over effects – is quite different from saying that it is likely to have an irreversible impact on a single important domain. It is a different claim again to suggest that AI may end up precipitating fundamental and unprecedented societal change on the level of the industrial revolution. In order to have productive discussions about the scale and nature of AI's potential societal impacts, it is helpful to unpack the key elements which distinguish these different types of societal transformation.

What is common to all these cases of societal change, we suggest, is that they lead to what we call practically irreversible change in trajectories of human life and progress. We follow Verbruggen 2013 in defining irreversible change as that which has long-lasting effects, is impossible (or extremely costly) to revoke, and creates, destroys or impairs something that has no functional equivalent. Here we talk of practically irreversible change since it is very difficult to be sure that a change is truly irreversible in theory; a societal change counts as practically irreversible if it cannot be reversed given the will, knowledge, and resources of any motivated group by whatever point in time that change can be said to exist or to have occurred[7]. Technologies may have practically irreversible societal impacts by enabling new products and capabilities that are so embedded in society our lives become dependent on them. For example, attempting to eliminate electricity at this point would very likely lead to catastrophic loss of life by disrupting key societal functions such as healthcare and international trade. Technologies may also have practically irreversible impacts by changing incentives and behavior in long-lasting and extremely difficult to change ways: for example, the invention and use of nuclear weapons fundamentally changed the calculus of great power conflicts. We therefore suggest that the notion of practically irreversible change should be considered core to what it means for change to be transformative on any level.

---

[6] Similar figures not using a logarithmic y-axis, show even more dramatic impact (e.g., see Clark 2001).
[7] This is to say that the change is no longer ongoing or an anticipated future event.

Closely related to the notion of practically irreversible change is the notion of "lock-in" of trajectories, either in the development of a technology (Arthur 1989) or in its use/impact on society (Wilson 2014). By lock-in we refer to the strong path dependence which emerges when some technology becomes so widely used for a certain application in society that it becomes extremely difficult to change paths (Shapiro et al. 1998). We are only concerned with lock-in, or path dependence, in the simplest sense, and without implications on the quality of the paths taken (Liebowitz and Margolis 1995). The example of the development and use of nuclear weapons is one example of technological lock-in: now that the world knows how to use this technology, warfare will never be the same. A more benign but perhaps still important example is the QWERTY keyboard, which was first patented in 1878 for use in typewriters and has been the most widely used keyboard for nearly 140 years (Noyes 1983). Although this keyboard is now commonly thought to be a suboptimal arrangement of keys, a number of downstream consequences of this original design mean that it is essentially now "locked-in" and very unlikely to change (David 1985).

The difference between electricity and nuclear weapons – the former being classed as a GPT where the latter is not – is in the *breadth* or generality of the societal change precipitated. While the invention of electricity irreversibly changed almost all aspects of life and society, nuclear power had impacts more contained to a few domains (*e.g.*, infrastructure and military). The transformative impacts of different technologies may also vary in their *extremity*: the magnitude of the changes in society they lead to[8]. While GPTs are defined as having widespread impact across life and society, their impact is not necessarily extreme enough to lead to notable changes in metrics of human progress and well-being[9], as we arguably saw as a result of the industrial revolution (Muehlhauser 2017).

To summarize: we suggest that what is fundamental to the notion of transformative change is that it constitutes **practically irreversible change** in certain trajectories of human life and progress. Beyond this, transformative change may be more or less **broad** in its impact - depending on the extent to which changes impact many different aspects of life and society - and more or less **extreme** - depending on the magnitude of the change relative to the time period. We believe that it would improve communication about the societal impacts of AI (and other technologies) if people were more explicit about these dimensions.

## 4. Defining Levels of Transformative AI

Based on the analysis above, we propose a definition of transformative AI which distinguishes between three levels[10]:

---

[8] Breadth and extremity are oftentimes correlated, and while there may exist examples for which this is not the case, we focus on examples where breadth and extremity vary together for pragmatic reasons.
[9] Metrics of human well-being might include subjective well-being, physical health, economic well-being, and social well-being (see Muehlhauser 2017 for further discussion). We also include economic growth as a measure of human progress in this category.
[10] Our proposed framework does not include some topics that may be considered components of societal transformation such as racial or gender-based biases embedded in computer vision and natural language processing. We note here that we are not ignoring these issues, but rather that they do not have the potential to lead to practically irreversible

**Table 3:** The Proposed Three Levels Defining Transformative AI

| Level of Transformativeness | Definition |
|---|---|
| **Narrowly Transformative AI** | Any AI technology or application with the potential to lead to practically irreversible change focused primarily in a specific domain or sector of society, such as warfare or education. Historical analogue: the impact of nuclear weapons on warfare. |
| **Transformative AI** | Any AI technology or application with potential to lead to practically irreversible change that is broad enough to impact most important aspects of life and society. One key indicator of this level of transformative change would be a pervasive increase in economic productivity (i.e., a 'productivity bonus'; Lipsey et al. 2005). Historical analogues: GPTs such as electricity and the internal combustion engine. |
| **Radically Transformative AI** | Any AI technology or application which meets the criteria for TAI, and with potential impacts that are extreme enough to result in radical changes to the metrics used to measure human progress and well-being, or to result in reversal of societal trends previously thought of as practically irreversible. This indicates a level of societal transformation equivalent to that of the agricultural or industrial revolutions. |

We emphasize the *potential* for societal impact when defining these levels of TAI, to acknowledge that we cannot say for certain whether a type of AI system will precipitate a given level of societal change in advance. Figure 2 below shows the three levels, historical examples for each level, and the advances in AI we suggest could lead to each level. While we do outline concrete examples below, our claim here is not necessarily that AI will inevitably progress through each of these levels. Rather, we suggest that using this delineation can reduce ambiguity in what it means to say AI will be "transformative", improving communication between AI researchers and policymakers about potential future impacts.

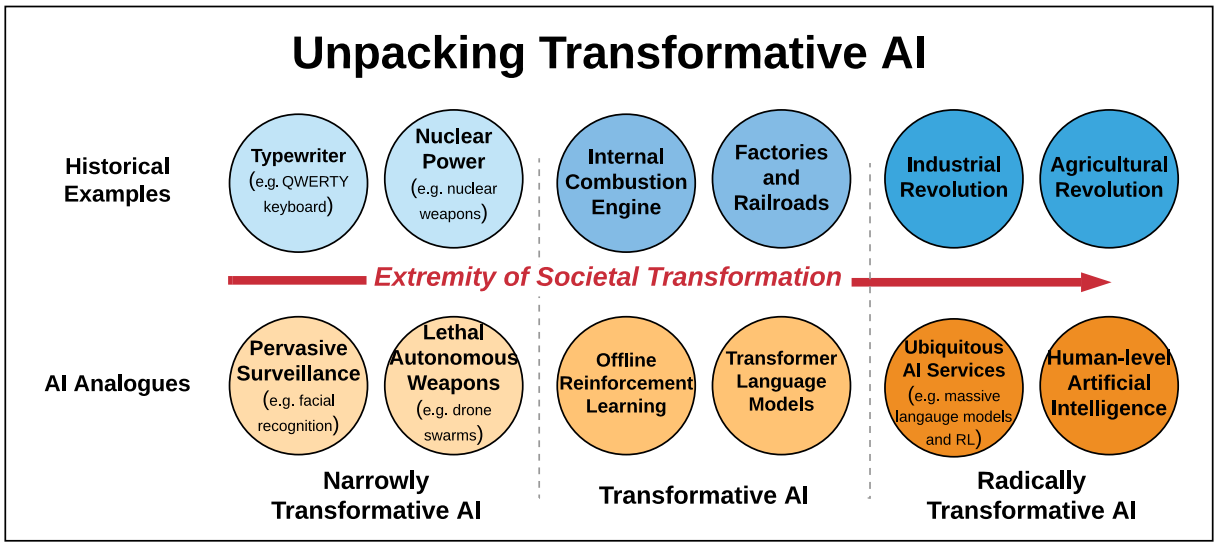

**Figure 2:** Our proposed levels of transformative AI and analogous AI technologies compared with historical examples of transformative GPTs.

**Plausible paths to the different levels**

change. Humanity has made much progress on these issues in the past century; thus, such issues are reversible to some significant degree.

It is plausible that we could see narrowly transformative impacts from more widespread use of current AI capabilities, without further technical advances. For example, widespread use of AI-driven surveillance technology could irreversibly change state powers, the nature of policing, and privacy, and lethal autonomous weapons such as drone swarms could irreversibly alter the nature of conflict in a way analogous to nuclear weapons.

Moving forward, scholars are suggesting that AI is the next major GPT, with potential to significantly impact economic growth and life across all sectors of society (Leung 2019; Brynjolfsson et al. 2018). We suggest two emerging AI technologies which could plausibly bring about the kind of broad societal impact associated with TAI[11]. First, most practical state-of-the-art applications of AI use supervised learning for pattern recognition or prediction tasks to aid in decision-making, but this approach of leveraging large amounts of data and powerful function approximation has not yet been able to yield the same real-world value with reinforcement learning. However, offline reinforcement learning holds enormous potential for the end-to-end automation of decision-making in a way that could transform domains from business to healthcare to robotics in ways that supervised learning alone will not (Levine et al. 2020). Recent work such as that on conservative Q-learning (Kumar et al. 2020) demonstrates that progress in a promising new class of novel solutions can mitigate the problem of distributional shift, which is the fundamental challenge precluding the practical, real-world application of large-scale, data-driven reinforcement learning.

Second, dramatic advances in natural language understanding over the past two years could plausibly usher in a long-anticipated era of practical human-machine interaction via language user interfaces (de Vries et al. 2020). These advances have been driven by transformer language models which have dominated general language understanding benchmarks. The most impressive work in this domain has demonstrated strong generalizability on language understanding tasks using few-shot learning (Brown et al. 2020), and a rigorous analysis suggests that continued scaling of transformer language models and dataset size will continue to lead to better performance with no plateau in sight (Kaplan et al. 2020). Such continued scaling of transformer language models or continued progress toward practical offline reinforcement learning offer evidence of plausible paths to the type of productivity bonus that would be associated with TAI.

The possible emergence of RTAI is of course more speculative but seems likely to evolve from the development of AI systems that can perform the majority of economically relevant tasks – whether in the form of many separate 'services' which collectively perform all tasks, or a single, human-level intelligence. It seems plausible that systems which seamlessly integrate grounded language user interfaces with powerful decision-making engines could, through a research and development process of human-in-the-loop recursive improvement (Drexler 2019), result in systems which can collectively replace the majority of current jobs. Further, current trajectories of progress in reinforcement learning[12] and multimodal models could result in systems capable of automating substantive parts of the scientific process, resulting in a speedup of scientific progress comparable to the industrial revolution (Karnofsky 2016). While it is beyond the scope of this article to assess the plausibility of these paths in detail, we believe these scenarios do demonstrate the plausibility of seeing RTAI without necessarily achieving fully human-level intelligence.

---

[11] These suggestions are intended to be examples rather than predictions. The second would likely be thought to be more plausible by AI researchers (see Bommasani et al. 2021).
[12] Silver et al. 2021 conjecture that reinforcement learning alone is sufficient for RTAI in the form of AGI.

## 5. Discussion

**Implications for the AI research community**

We hope this definition and analysis can prompt clearer and more substantive discussion and analysis around what kinds of AI technologies are likely to lead to different levels of societal impact. In distinguishing three levels of transformative AI, we highlight how AI systems may have profound and long-lasting impacts on society well before anything close to 'human-level intelligence' is reached. Current AI capabilities, if deployed more widely, could irreversibly change key domains in 'narrow' but long-lasting ways. Current avenues of AI progress, such as those in offline reinforcement learning and transformer language models, could lead to widespread societal transformation on the level of previous GPTs. We also believe these distinctions can help prompt more substantive discussion and analysis around which kinds of advances in AI are likely to lead to different levels of societal impact – e.g., what does it mean to be "more profound than fire or electricity?" (Pichai and Schwab 2020).

In distinguishing between NTAI and TAI we highlight that existing or near-future AI systems have the potential to transform society in the narrow sense of precipitating practically irreversible change to important domains. The distinction between TAI and RTAI in turn emphasizes that we may see widespread societal transformation well before AI systems achieve fully general or human-level capabilities, or before we see societal transformation on the level of the agricultural or industrial revolutions. For example, if advances were made in offline reinforcement learning that enabled powerful deep reinforcement learning techniques to leverage the values of big data, we could see a pervasive economic impact[13]. We believe that such transformative impacts, comparable to those from previous GPTs, are currently a neglected topic in discussion of the societal impacts of AI. Moreover, since many believe RTAI to be hundreds or thousands of years away, or even impossible (Grace et al. 2018), the notion of TAI offers a more widely acceptable alternative for a broader community of scholars to discuss potential extreme societal impacts of advanced AI.

Our analysis opens up many questions for future discussion and debate, including what kinds of AI technologies have potential to lead to different levels of transformation. One might argue that widespread use of AI-driven surveillance technology or lethal autonomous weapons could have much wider-ranging impacts on life and society than we suggest here, if for example the former resulted in robust authoritarianism or the latter to unprecedented great power conflict (Dafoe 2018). Certainly, to suggest that the sole impact of widely practical offline reinforcement learning would be a 'productivity bonus' is an oversimplification – the resulting potential impacts on all other areas of life and society warrant much further investigation and may be at least as important to prepare for as the economic impacts. Some may challenge whether even human-level AI could alone have an impact on the same level as the industrial revolution, or conversely, argue that it is reasonable to expect we would see such impacts before achieving fully general or human-level AI.

We note that the category of RTAI stands out as being understood by analogy to the agricultural and industrial revolutions, but not with any single technology as a historical analogue. The invention of steam power and its subsequent impacts on factories and railroads might be a candidate for such a technology in the case of the industrial revolution, but the singular value of the steam engine is not widely agreed upon in the literature. Indeed, there is no clear evidence or consensus that any single technology has alone precipitated change on the level we are describing as "radical societal transformation" - historically these changes seem to have resulted from clusters of technologies potentially in interaction with other societal factors. However, AI is arguably unique in that it does not necessarily represent a single technology, but an underlying method leading to a cluster of different technologies: including, for example, natural language processing, computer vision, and robotic learning. Thus, while it is plausible that a single AI technology, such as HLAI, could lead to RTAI, it is also possible that a cluster of different AI technologies could lead to TAI or RTAI.

---

[13] Economists generally expect the broader economic impact of future AI systems to come from the automation of human labor (Frank et al. 2019; Aghion et al. 2017), with some believing it may lead to explosive economic growth (Nordhaus 2021; Hanson 2001).

We believe that transformative societal impacts from TAI and RTAI warrant further consideration within the AI research community. AI researchers and engineers are beginning to take seriously the potential risks and harms associated with specific applications of AI, such as in the military, and this has begun to impact their work, as demonstrated for example by the Google employee backlash over project Maven (Shane and Wakabayashi 2018). The societal risks associated with TAI and RTAI will be much more severe – for example, decision-making engines could be used for centralized economic planning which, depending on numerous factors, could be very positive or very negative (Parson 2020). In 2020, the NeurIPS conference, the premier conference in AI and machine learning, began requiring submissions to address the broader impacts of their research, and the NeurIPS 2021 concluded that this was an important step forward that led to constructive steps toward more responsible machine learning research (Beygelzimer et al. 2021). An important next step will be to begin considering how broader trends in societal applications of AI and in avenues of progress may impact key domains of society in longer-lasting and less easily reversible ways, beyond the impacts of any specific paper. We believe that doing this effectively will require greater interdisciplinary collaboration between experts in AI, social science, and policy. The positive progress of the NeurIPS broader impact statement is a good example of the value that such collaboration can add. To these ends we recommend that, for example, major AI conferences consider hosting workshops or discussion sessions aimed at bringing together different areas of expertise to reflect on the priorities of the AI field and their possible impacts, and that funding bodies consider awarding grants to interdisciplinary projects aiming to rigorously explore the future impacts of AI progress. Individual researchers and groups might also contribute to these efforts by hosting interdisciplinary discussion groups around these questions of how we might see different types of transformative change from AI in future.

**Implications for futures researchers and practitioners**

The proposed framework is directly intended to shift anticipatory assumptions about the future regarding the role of AI (e.g., that radical societal transformation from AI can only come from anthropomorphic notions of AGI such as HLMI), with the objective of opening up new pathways to futures involving the development of safe and beneficial RTAI. This goal is important for both the AI research community as well as futures researchers and practitioners, but we feel that it is most salient to those at the intersection of these two groups. In particular, we hope this framework can be used as a starting point for broader exploration of a range of future scenarios involving AI.

Making overly-narrow assumptions about possible AI futures is especially dangerous, given the potentially transformative and irreversible societal impacts AI may have. Though people may disagree about how dangerous advanced AI is likely to be, we refer back to Sundar Pichai's quote from the beginning of the article (Pichai and Schwab 2019). It is undoubtedly significant that the chief executive officer of the world's largest AI firm does not attempt to minimize or obfuscate the real risks of negative outcomes from AI, when speaking to participants of one of the most elite annual gatherings of world leaders in modern history. Thinking clearly and comprehensively about the possible risks from AI is essential to ensuring safe and beneficial outcomes, and this framework should aid futures researchers and practitioners in doing so.

In general, there have been substantial efforts in the literature which have attempted to forecast the arrival of notions of AGI such as HLMI (Grace et al. 2018; Müller and Bostrom 2016; Baum et al. 2011). There have also been efforts to explore plausible futures involving similarly advanced AI (Tegmark 2017; Bostrom 2014; Kurzweil 2005), but few of these have used rigorous futures studies methods. We feel that this presents an opportunity for futures researchers and practitioners to use techniques such as the Delphi and scenario planning to explore plausible futures that TAI and RTAI could bring. Due to the entangled nature of the wicked problems challenging the development of safe and beneficial AI (Author 2018), novel techniques have been proposed combining futures methods (Author 2019). We believe that novel futures methodologies such as this (e.g., combining techniques like scenario mapping and the Delphi) could be very useful for exploring many of the questions mentioned in this section.

**Future directions for research**

*What types of AI developments could lead to each level of societal change?*

This analysis offers many paths for future discussion and debate. For example, would the widespread adoption and application of current AI technologies lead to societal impacts that would constitute TAI? One might argue that the ubiquitous use of machine learning algorithms could have could have such impacts, if it led to practically irreversible change that was broad enough to impact most important aspects of life and society. How might widespread use of language user interfaces or decision-making engines affect society beyond purely economic impacts? Of course, to suggest the sole impact would be a 'productivity bonus' is an oversimplification – the impact on politics, power, and people's daily lives are at least as important to prepare for and warrant further exploration.

It would also be valuable to develop a more rigorous analytical framework for making and assessing claims about which AI developments may lead to different levels of societal transformation, given such claims are likely to be highly subjective and uncertain. Useful methodologies here might include those from forecasting and foresight to explore possible developments in AI technologies and their impacts, and aggregation of expert opinion in particular to synthesize diverse perspectives and expertise on these questions.

*How do different levels of societal transformation relate to each other?*

In both the case of the agricultural and industrial revolutions, it appears that radically transformative societal change was at least initially driven by advances in a single critical technology (*i.e.*, a GPT): the domestication of plants in the case of the agricultural revolution, and steam power in the case of the industrial revolution. By contrast, the invention of nuclear weapons and the transformative change that followed did not seem to later lead to more broadly transformative technologies, at least not directly. It is possible that in some cases, a new technology might lead to lower levels of societal change without later developing in ways that lead to higher levels of societal transformation; and vice versa, that radical innovation (*i.e.*, discontinuous technological progress) could lead to radically transformative impacts *without* any lower-level impacts serving as warning signs.

The question of when and whether we should expect lower levels of transformative change to precede or directly lead to higher ones is important for AI, and is currently underexplored. In particular, if RTAI may emerge without being preceded by incrementally more transformative AI, the work needed to prepare for its impacts will look quite different from a scenario where we have more warning signs. One way to explore the relationship between TAI and RTAI scenarios in more depth would be to look at how various transformative technologies have ended up precipitating radical societal change historically: is it possible to better understand why steam power ended up precipitating societal change on a different scale to electricity, for example? Given current uncertainty about this question, we suggest that various scenarios in which RTAI may emerge deserve preparation and attention.

*Over what timeframe could transformative impacts of AI occur?*

As AI is integrated with so many existing information technologies, it seems plausible that it could lead to a level of societal transformation similar to that of other GPTs such as electricity, in a much shorter period of time. The examples we proposed of advanced offline reinforcement learning (Levine et al. 2020) and scaled transformer language models (Brown et al. 2020; Kaplan et al. 2020) are plausible relatively quick paths to TAI. Such a rapid rise of TAI could create problems for organizations and policy makers that previous GPTs or technological transformations have not, even if the impact is not on the level of RTAI. For example, a rapid rise to TAI may make it difficult for entrepreneurs to deploy existing labor in new ways as they have when previous GPTs have led to automation and labor demand (Brynjolfsson et al. 2018). Further research exploring arguments and analysis for TAI developments arising on different timeframes, and what the impacts of a particularly rapid rise to TAI might look like, would therefore be particularly valuable.

*What are other implications for the framework?*

There are numerous other implications of the framework regarding future work. One interesting consideration involves the fact that advanced AI, such as TAI or RTAI, offers an opportunity to, if managed appropriately, affect change that could have previously been considered practically irreversible. For example, the effects of colonialism could be argued to have been locked in for many nations to some degree for a long time, and the prospect of achieving economic equality may have been thought virtually impossible until recently. However, RTAI may easily reverse entrenched inequalities remnant from colonialism, but it carries the potential to both erase inequalities or to strengthen them. This is a unique and interesting problem that is highly suitable for future work.

*How can economics help to better understand transformative and radically transformative change from AI?*

The impacts of advanced AI, such as TAI and RTAI, have long been a subject of interest to economists (Simon 1965; Hanson 2001), and our proposed framework is clearly closely tied to economic productivity. Consequently, there exists substantial potential for future work to explore these ties more rigorously, in order to more objectively define and understand the proposed dimensions of breadth and extremity. Recent work has explored economic growth for robots (Acemoglu and Restrepo 2020), AI (Aghion et al. 2017) and even for a singularity (i.e., one notion for how RTAI may come to be; Nordhaus 2021). Such studies explore the substitutability of conventional factors of production (e.g., non-AI capital, labor) with AI or robotics, and it would be very beneficial if future work could explore the degree of substitutability between factors of production that could be expected with differing levels of advanced AI such as those proposed here.

## 6. Conclusion

We have examined existing literature to frame the transformative potential of AI relative to impacts of historical technologies. The analogies presented here are intended to help convey to readers three significantly different levels of possible societal transformation from AI. We suggest that the possible emergence of TAI – AI technologies or applications with potential to lead to practically irreversible societal and economic change across all of society – is currently a particularly neglected topic, since existing discussions tend to focus either on immediate impacts of AI or the extreme possibility of human-level or superintelligent AI (Author et al. 2020). It seems plausible that TAI will arise over the next decade (Grace et al. 2018), possibly through emerging AI technologies such as advanced offline reinforcement learning and scaled transformer language models. Currently nations are not prepared for this, and without dramatic action from policy makers the anticipated arrival of TAI could have severe consequences for much of the world's population. The levels proposed in this paper give researchers, strategic planners and decision makers a more effective framework through which they can understand possible futures involving advanced AI, to prepare for the impacts of different levels of societal transformation from AI, and to allocate resources accordingly. We suggest that, due to the potential for rapid TAI development, future work should urgently explore plausible paths to TAI and their consequences.

# Acknowledgements

We would like to thank Allan Dafoe, Ben Garfinkel, Matthijs Maas, Alexis Carlier, David Manheim, Shahar Edgerton Avin and Jose Hernandez-Orallo for their comments and discussion at different stages of this project. This collaboration was made possible by funding from the Berkeley Existential Risk Initiative.